\documentclass[a4paper, accepted=2022-05-31]{quantumarticle}
\pdfoutput=1
\usepackage{amsmath}
\usepackage{amsthm}
\usepackage[numbers]{natbib}
\usepackage{hyperref}
\usepackage{amssymb}
\usepackage{graphicx}
\usepackage{braket}
\usepackage{bbm}
\usepackage{bm}
\DeclareMathOperator{\id}{\mathbbm{1}}
\newcommand{\beq}{\begin{equation}}
\newcommand{\eeq}{\end{equation}}

\begin{document}
\title{Universal quantum circuits for quantum chemistry}
\author{Juan Miguel Arrazola}
\email{juanmiguel@xanadu.ai}
\affiliation{Xanadu, Toronto, ON, M5G 2C8, Canada}
\author{Olivia Di Matteo}
\affiliation{Xanadu, Toronto, ON, M5G 2C8, Canada}
\author{Nicol\'as Quesada}
\affiliation{Xanadu, Toronto, ON, M5G 2C8, Canada}
\author{Soran Jahangiri}
\affiliation{Xanadu, Toronto, ON, M5G 2C8, Canada}
\author{Alain Delgado}
\affiliation{Xanadu, Toronto, ON, M5G 2C8, Canada}
\author{Nathan Killoran}
\affiliation{Xanadu, Toronto, ON, M5G 2C8, Canada}

\begin{abstract}
Universal gate sets for quantum computing have been known and studied for decades, yet less is understood about universal gate sets for particle-conserving unitaries, which are operations of interest in quantum chemistry. 
In this work, we show that controlled single-excitation gates in the form of Givens rotations
are universal for particle-conserving unitaries. Single-excitation gates describe an arbitrary $U(2)$ rotation on the two-qubit subspace spanned by the states $|01\rangle, |10\rangle$, while leaving other states unchanged -- a transformation that is analogous to a single-qubit rotation on a dual-rail qubit. The proof is constructive, so our result also provides an explicit method for compiling arbitrary particle-conserving unitaries. Additionally, we describe a method for using controlled single-excitation gates to prepare an arbitrary state of a fixed number of particles. We derive analytical gradient formulas for Givens rotations as well as decompositions into single-qubit and CNOT gates. Our results offer a unifying framework for quantum computational chemistry where every algorithm is a unique recipe built from the same universal ingredients: Givens rotations.
\end{abstract} 
\maketitle
\section{Introduction}
Quantum algorithms for quantum chemistry rely on the ability to prepare states
that represent fermionic wavefunctions~\cite{mcardle2020quantum, cao2019quantum,
  reiher2017elucidating}. These can correspond to ground and excited states of
molecular Hamiltonians, which can then be employed to compute properties of the
molecule~\cite{peruzzo2014variational, o2016scalable, mcclean2017hybrid,
  ryabinkin2018constrained, mitarai2020theory, ibe2022calculating}. In most
  molecules and materials, the number of particles is a conserved quantity and
  quantum states that do not respect this condition are unphysical. Valid
  quantum states thus occupy only a subspace of the available Hilbert
  space. Algorithms that access subspaces of different
  particle number are therefore not only wasteful, but can potentially lead to
  incorrect outcomes. This motivates the use of gate sets that preserve
  subspaces of fixed particle number. We focus on the Jordan-Wigner
representation~\cite{jordan1928paulische}, which encodes the subspace of states with $k$
particles in $n$ spin-orbitals into $n$ qubits. This space is spanned by the set of all $n$-qubit states with Hamming weight $k$, i.e., states with $k$ ones and $n-k$ zeros. To ensure that output states remain valid, quantum circuits for quantum chemistry benefit from employing gates that preserve Hamming weight and therefore particle number.

Universal gate sets capable of synthesizing arbitrary unitary operations have been known for decades~\cite{lloyd1995almost, divincenzo1995two}. Famously, the set of arbitrary single-qubit rotations and CNOT gates is universal for quantum computation~\cite{nielsen2000quantum}. Yet less is known about universal gate sets for particle-conserving unitaries, which are precisely the operations of interest in quantum chemistry. A notable exception is the work of Ref.~\cite{oszmaniec2017universal}, which provides a non-constructive proof that Gaussian fermionic operations together with a non-Gaussian gate is universal for particle-conserving transformations. Additionally, there are several proposals for preparing states of fermionic systems~\cite{ward2009preparation, wang2009efficient, jiang2018quantum}, some of which are designed to preserve particle number and other symmetries~\cite{yordanov2020efficient, gard2020efficient, anselmetti2021local}. A universal set of particle-conserving gates, together with a constructive method for compiling arbitrary unitaries, would constitute a flexible and composable framework for designing arbitrary quantum circuits for quantum chemistry.

Currently, a wide variety of quantum circuit architectures have been proposed to
prepare states of many-body fermionic systems, particularly in the context of
variational quantum algorithms. These include chemically-inspired
circuits~\cite{yung2014transistor, romero2018strategies}, adaptive
circuits~\cite{grimsley2019adaptive, tang2021qubit}, hardware-efficient
circuits~\cite{kandala2017hardware,barkoutsos2018quantum}, and other specialized
methods~\cite{ryabinkin2018qubit, matsuzawa2020jastrow}. This situation is not
ideal because quantum algorithm developers are seemingly faced with a choice among different proposals rather than having access to universal building blocks to construct any desired operation. There is therefore a need for a unified conceptual framework for constructing and designing quantum circuits for quantum chemistry. This is particularly important for software implementations that aim to provide users with maximum flexibility and to be adaptable in incorporating future algorithmic innovations.

In this work, \emph{we provide such a framework} by giving a constructive proof
that controlled single-excitation gates are universal for particle-conserving
unitaries.~A single-excitation gate performs an arbitrary $U(2)$ transformation
in the subspace $\ket{01}, \ket{10}$ while leaving other basis states unchanged:
\beq\label{eq:single-excitation}
\begin{pmatrix}
1 & 0 & 0 & 0\\
0 & a & c & 0\\
0 & b & d & 0\\
0 & 0 & 0 & 1
\end{pmatrix},
\eeq
where $U=\begin{pmatrix} a & c\\ b & d\end{pmatrix}$ is a general $2\times2$ unitary. Single-excitation gates can be viewed as an extension of \emph{Givens rotations} to unitary two-dimensional transformations. A controlled single-excitation gate, which applies this Givens rotation depending on the state of a third qubit, can be described by the unitary
\beq\label{eq:control-single-excitation}
\begin{pmatrix}
1 & 0 & 0 & 0 & 0 & 0 & 0 & 0\\
0 & 1 & 0 & 0 & 0 & 0 & 0 & 0\\
0 & 0 & 1 & 0 & 0 & 0 & 0 & 0\\
0 & 0 & 0 & 1 & 0 & 0 & 0 & 0\\
0 & 0 & 0 & 0 & 1 & 0 & 0 & 0\\
0 & 0 & 0 & 0 & 0 & a & c & 0\\
0 & 0 & 0 & 0 & 0 & b & d & 0\\
0 & 0 & 0 & 0 & 0 & 0 & 0 & 1
\end{pmatrix}.
\eeq

In addition to the universality result, we propose an explicit algorithm using excitation gates to prepare an arbitrary state with a fixed number of particles. We derive analytical gradient formulas for Givens rotations and argue that they are ideal building blocks in variational quantum circuits for quantum chemistry. Our results and framework allow the design of quantum algorithms for quantum chemistry from a universal gate set of operations that respect the symmetries of  fermionic systems and have a transparent physical interpretation. Although we focus on the specific case of quantum chemistry, our results apply more generally to the simulation of many-body fermionic systems in the Jordan-Wigner representation. As described in Ref.~\cite{arrazola2021differentiable}, the framework described in this work has been implemented in the open-source library PennyLane, providing actual building blocks for building arbitrary particle-conserving circuits.

The rest of this manuscript is organized as follows. We introduce the basic concepts of particle-conserving unitaries and Givens rotations in Sec.~\ref{sec:universal}. We then show in Sec.~\ref{sec:proof} that controlled single-excitation gates are universal for particle-conserving unitaries. In Sec.~\ref{sec:state-prep} we describe a universal method for preparing states with a fixed number of particles. We discuss the role of Givens rotations in variational quantum circuits in Sec.~\ref{sec:variational} and conclude in Sec.~\ref{sec:conclusion}.

\section{Universal gate set}\label{sec:universal}
We introduce basic concepts and notation that are relevant before presenting the main results. For simplicity and generality, we employ abstract notions of particles and operations, without establishing an explicit connection to concepts such as electrons, spin-orbitals, or fermionic operators.

\subsection{Particle-conserving unitaries}

Define the qubit ladder operators $\sigma^\dagger = (X +iY)/2$, $\sigma = (X -iY)/2$, where $X,Y$ are Pauli matrices, and the total number operator
\beq
N = \sum_i \sigma^\dagger_i\sigma_i.
\eeq
For a computational basis state $\ket{x}$ it holds that $N\ket{x}=w(x)\ket{x}$, where $w(x)$ is the Hamming weight of the bit string $x$. We define the Hamming weight to be equal to the number of particles and refer to eigenstates of the total number operator as states with a fixed number of particles. A unitary gate $U$ is deemed particle-conserving if 
\beq
[U, N] = 0.
\eeq
A particle-conserving unitary maps states with a fixed number of particles to
other states with the same fixed number of particles. Any product of particle-conserving unitaries is also particle-conserving, so any quantum circuit composed of particle-conserving gates is guaranteed to perform a particle-conserving transformation.

\begin{figure}[t]
\includegraphics[width=0.93\columnwidth]{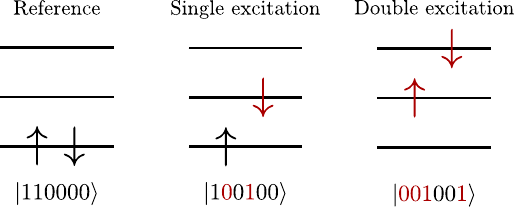}
\caption{Jordan-Wigner representation of states with a fixed number of particles. Each qubit corresponds to an orbital with a specific spin orientation. The state of the qubit determines whether that spin-orbital is occupied or not. All basis states can be obtained from a reference state, in this case $\ket{110000}$, by a specific excitation. For instance, the state $\ket{100100}$ is obtained by exciting a particle from qubit 2 to 4, while the state $\ket{001001}$ is obtained by exciting both particles to qubits 3 and 4.}\label{fig:states}
\end{figure}

The space of all states with $k$ particles on $n$ qubits, denoted as
$\mathcal{H}_k$, is spanned by the set of computational basis states with
Hamming weight $k$. In general, any state of a fixed number of particles can be
interpreted as an excitation from a reference state. Unless stated otherwise, we
use the state $\ket{11\cdots 100\cdots0}$ with all particles in the first $k$
qubits as the reference state. This is illustrated in Fig~\ref{fig:states}. We
use the Hamming distance $\sum_{i=1}^n x_i\oplus y_i$ to denote the number of
qubits where the computational basis states $\ket{x}$ and $\ket{y}$ differ. For
states with an equal number of particles, the Hamming distance is an even
number. An example of particle-conserving unitaries are fermionic
  linear-optical transformations as explored in
  Ref.~\cite{oszmaniec2017universal}, which together with a non-Gaussian gate
  form a universal gate set for particle-conserving unitaries.

Two states $\ket{x}$ and $\ket{y}$ having an equal number of particles are said to differ by an excitation of order $\ell$ if their Hamming distance is equal to $2\ell$. For example, the states $\ket{1100}$ and $\ket{0101}$ differ by a single excitation (order 1) from the first to the fourth qubit. Similarly, the state $\ket{0011}$ differs from $\ket{1100}$ by a double excitation (order 2).  

Any particle-conserving unitary acting on states with $k$ particles on $n$ qubits can be represented as a block-diagonal unitary performing a general $U(d)$ transformation on the subspace $\mathcal{H}_k$ with dimension $d=\text{dim}(\mathcal{H}_k)$. For universality, it is therefore sufficient to consider a set of particle-conserving gates that is universal for the subspace 
$\mathcal{H}_k$.

\subsection{Givens rotations}
As discussed above, any two states with a fixed number of particles differ by an excitation of a given order. It is therefore convenient to work with a set of quantum gates that create superpositions between the original and the excited state. In the simplest non-trivial case of a single particle and two qubits, these correspond to gates that perform arbitrary $U(2)$ rotations between the states $\ket{10}, \ket{01}$ while leaving other basis states unchanged. For example, restricting to the case where the gate has only real parameters, a two-qubit particle-conserving unitary can be written as
\beq
G(\theta)=\begin{pmatrix}
1 & 0 & 0 & 0\\
0 & \cos (\theta) & -\sin (\theta) & 0\\
0 & \sin(\theta) & \cos(\theta) & 0\\
0 & 0 & 0 & 1
\end{pmatrix},\label{eq:givens}
\eeq
where we use the ordering $\ket{00}, \ket{01}, \ket{10}, \ket{11}$ of two-qubit computational basis states.

This is an example of a \emph{Givens rotation}: a rotation in a two-dimensional subspace of a larger space. In this case, the Givens rotation acts as a two-qubit \emph{single-excitation gate}, coupling states that differ by a single excitation. More generally, we can extend the concept of Givens rotations to $U(2)$ transformations in two-dimensional subspaces, where a general single-excitation gate can be written as
\beq
G = \begin{pmatrix}
1 & 0 & 0 & 0\\
0 & a & c & 0\\
0 & b & d & 0\\
0 & 0 & 0 & 1
\end{pmatrix},
\eeq
where $|a|^2+|c|^2=|b|^2+|d|^2=1$ and $ab^*+cd^*=0$ to ensure unitarity. 
We can also consider four-qubit \emph{double-excitation gates} $G^{(2)}$ which perform a general $U(2)$ rotation on the subspace spanned by the states $\ket{0011}, \ket{1100}$
\begin{align}
G^{(2)}\ket{0011} &= a\ket{0011} + b\ket{1100},\\
G^{(2)}\ket{1100} &= d\ket{1100} + c\ket{0011},
\end{align}
while leaving all remaining four-qubit states unchanged. Double-excitation gates can also perform rotations in two-dimensional subspaces defined by pairs of four-qubit states with Hamming distance four, namely $\ket{1010}, \ket{0101}$ and $\ket{1001}, \ket{0110}$.

We can generalize to excitation gates of order $\ell$. These are unitary Givens rotations acting on the space of $2\ell$ qubits that couple the states $\ket{\bm{1}_\ell\bm{0}_\ell}:=\ket{1}^{\otimes \ell}\ket{0}^{\otimes \ell}$ and $\ket{\bm{0}_\ell\bm{1}_\ell}:=\ket{0}^{\otimes \ell}\ket{1}^{\otimes \ell}$ as
\begin{align}
G^{(\ell)} \ket{\bm{0}_\ell\bm{1}_\ell} &= a\ket{\bm{0}_\ell\bm{1}_\ell} + b\ket{\bm{1}_\ell\bm{0}_\ell},\label{eq:g^l_1}\\
G^{(\ell)}\ket{\bm{1}_\ell\bm{0}_\ell} &= d\ket{\bm{1}_\ell\bm{0}_\ell} +c\ket{\bm{0}_\ell\bm{1}_\ell},\label{eq:g^l_2}
\end{align}
while acting as the identity on all other states. Similar Givens rotations can be defined for permutations of the states $\ket{\bm{1}_\ell\bm{0}_\ell}, \ket{\bm{0}_\ell\bm{1}_\ell}$, i.e., excitation gates of order $\ell$ also include rotations on all pairs of states of $2\ell$ qubits with Hamming distance $2\ell$. By construction, these excitation gates are particle-conserving since they only couple states with an equal number of particles. 

As explained in more detail in Sec.~\ref{sec:proof}, on their own, excitation gates as Givens rotations are not universal for
  particle-conserving unitaries. We thus consider \emph{controlled} excitation gates, which apply an excitation gate depending on the state of a control qubit. In particular, we focus on the three-qubit controlled single-excitation gate
\beq
CG = \begin{pmatrix}
1 & 0 & 0 & 0 & 0 & 0 & 0 & 0\\
0 & 1 & 0 & 0 & 0 & 0 & 0 & 0\\
0 & 0 & 1 & 0 & 0 & 0 & 0 & 0\\
0 & 0 & 0 & 1 & 0 & 0 & 0 & 0\\
0 & 0 & 0 & 0 & 1 & 0 & 0 & 0\\
0 & 0 & 0 & 0 & 0 & a & c & 0\\
0 & 0 & 0 & 0 & 0 & b & d & 0\\
0 & 0 & 0 & 0 & 0 & 0 & 0 & 1
\end{pmatrix}.
\eeq
A particular example of a controlled single-excitation gate is the controlled
SWAP, or Fredkin gate
\beq
F = \begin{pmatrix}
1 & 0 & 0 & 0 & 0 & 0 & 0 & 0\\
0 & 1 & 0 & 0 & 0 & 0 & 0 & 0\\
0 & 0 & 1 & 0 & 0 & 0 & 0 & 0\\
0 & 0 & 0 & 1 & 0 & 0 & 0 & 0\\
0 & 0 & 0 & 0 & 1 & 0 & 0 & 0\\
0 & 0 & 0 & 0 & 0 & 0 & 1 & 0\\
0 & 0 & 0 & 0 & 0 & 1 & 0 & 0\\
0 & 0 & 0 & 0 & 0 & 0 & 0 & 1
\end{pmatrix}.
\eeq

\begin{figure}[t]
\centering
\includegraphics[width=0.63\columnwidth]{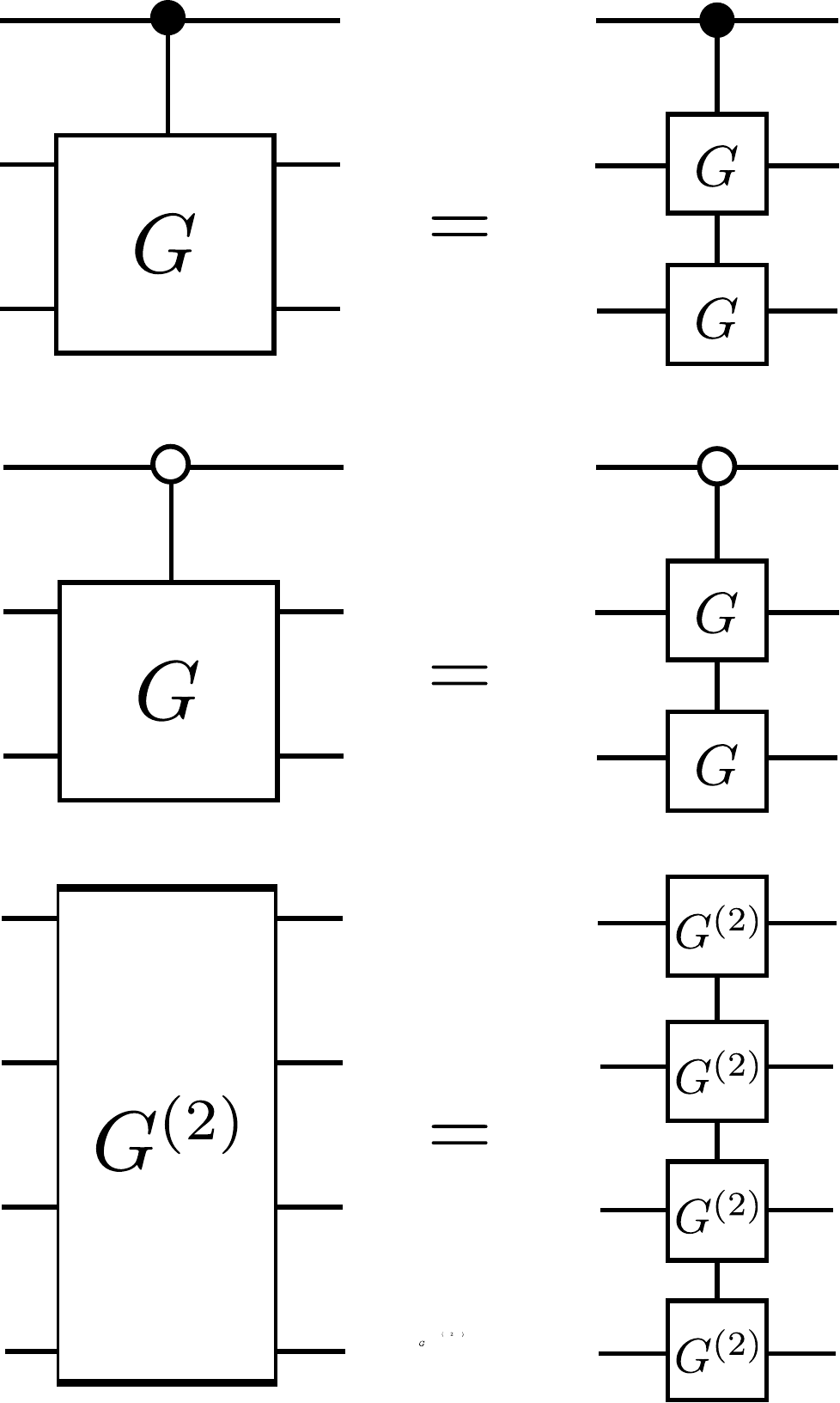}
\caption{The three-qubit controlled single-excitation gate, which is universal for
  particle-conserving unitaries, and the four qubit double-excitation gate. The controlled single-excitation gate on the top denotes a control on
  state $\ket{1}$ while the one below shows a control on state $\ket{0}$. The figures on the right are alternative representations of the excitation gates, which are useful for drawing circuits when the qubits are not adjacent. The bottom circuit is a double-excitation gate.}\label{fig:excitation-gates}
\end{figure}

Controlled gates usually refer to the case where a gate is applied only if the
control qubit is in state $\ket{1}$. Throughout this work, we more generally use
the term controlled gate to include also the case where gates are applied only
if the control qubit is in state $\ket{0}$. All such controlled gates are
particle-conserving. Controlled single-excitation gates and double-excitation
gates are illustrated in Fig. ~\ref{fig:excitation-gates}. It is also useful
  to note that Givens rotations can also be used to enforce spin conservation by
  coupling only qubits corresponding to spin-orbitals with the same spin
  projection. This allows a straightforward method for ensuring that spin-conservation is also respected in a quantum circuit. 

It is helpful to contrast Givens rotations to fermionic excitations, which
  are operators of the form $\exp[-i\theta(a_i^\dagger a_j-a_j^\dagger a_i)]$
  for a single excitation, where $a,a^\dagger$ are fermionic ladder
  operators. Higher-order excitations are constructed with higher-order
  monomials of the ladder operators. Fermionic excitations are expressed as
  qubit operators through the Jordan-Wigner mapping $a^\dagger_i\rightarrow
  \sigma^+_i\prod_{j<i}Z_j$, where $\sigma^{\pm}=(X\pm iY)$. Fermionic
excitations also couple subspaces of equal particle number, but they are more
difficult to analyze, implement, and simulate. The universality proof of the
next section demonstrates that it is possible to instead employ Givens rotations
directly since, as we will show later in Sec.~\ref{sec:state-prep}, they can be used to prepare any desired state. 

\section{Proof of Universality}\label{sec:proof}
In this section, we show that controlled single-excitation gates are universal for particle-conserving unitaries. We use standard textbook methods similar to those used for proving the universality of single-qubit and CNOT gates~\cite{nielsen2000quantum}. The proof follows these main steps: 
\begin{enumerate} 
\item For particle-conserving unitaries, we show that the relevant $U(2)$ transformations are excitation gates controlled on multiple qubits. Using the established result that $U(d)$ transformations can be decomposed into products of $U(2)$ transformations, it follows that excitation gates controlled on multiple qubits are universal for particle-conserving unitaries.
\item We show that any excitation gate controlled on multiple qubits can be decomposed in terms of multiply-controlled single-excitation gates.
\item Finally, we show that single excitation gates controlled on multiple qubits can be decomposed into three-qubit controlled single-excitation gates, which are therefore also universal.
\end{enumerate}

The main insight of this construction is that single-excitation gates are
analogous to single-qubit gates. Indeed, the subspace $\ket{10}, \ket{01}$ can be
interpreted as a dual-rail encoding of a single qubit. This allows standard
methods to carry over to the particle-conserving case, with few
modifications. Fredkin gates, which are a specific type of control
single-excitation gate, play a special role: they are used to extend controlled
single-excitation gates to controlled gates over multiple qubits. Fredkin gates have been shown to be universal for reversible computations in dual-rail encodings~\cite{aaronson2015classification}. \\

\subsection{Excitation gates with multiple controls}
It is well-established that any $U(d)$ transformation can be decomposed into a product of $U(2)$ transformations acting on arbitrary two-dimensional subspaces~\cite{reck1994experimental, nielsen2000quantum, clements2016optimal, de2018simple}. As discussed in the previous section, any state of a fixed number of particles can be obtained by applying an excitation to a reference state. This result holds more generally: any two states of a fixed number of particles differ by a specific excitation. \\

Consider two $k$-particle states on $n$ qubits $\ket{x},\ket{y}$, with Hamming
distance $2\ell$. Without loss of generality, since this can be guaranteed by
relabelling, suppose the first $k-\ell$ qubits are set to $1$ for both states
and the last $n-k-\ell$ qubits are set to $0$. The remaining $2\ell$ qubits are
in different states, meaning that they can be mapped to each other by exciting
the particles from the $\ell$ occupied qubits to the $\ell$ unoccupied ones. 

\begin{figure}[t]
\centering
\includegraphics[width=0.75\columnwidth]{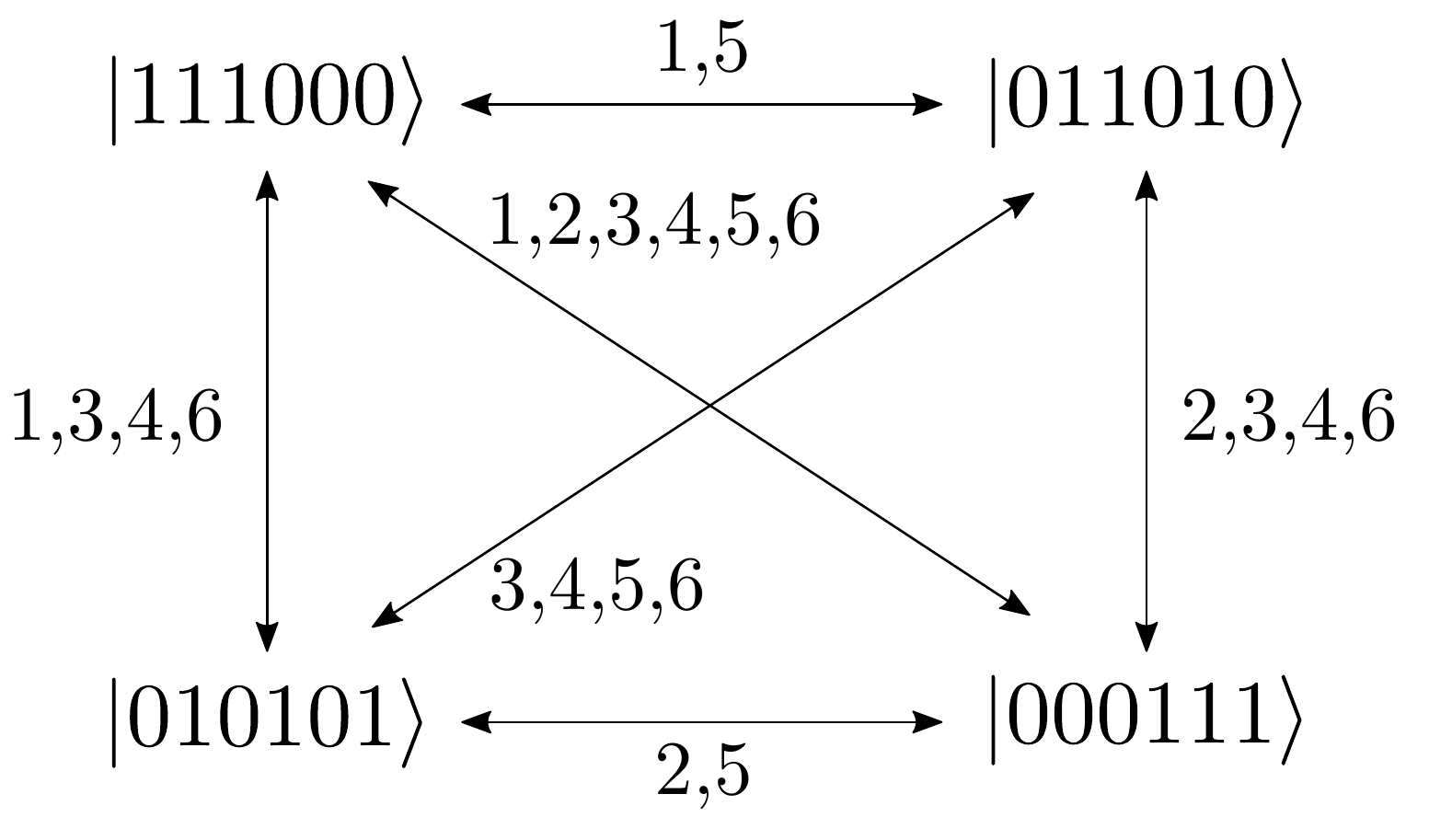}
\caption{Any pair of states with a fixed number of particles can be related by an excitation. The states $\ket{111000}$ and $\ket{011010}$ differ by a single excitation between qubits 1 and 5. The states $\ket{011010}$ and $\ket{000111}$ differ by a double excitation between qubits 2,3,4,6, while $\ket{000111}$ and $\ket{111000}$ are connected by a triple excitation acting on all qubits.}\label{fig:excitation_links}
\end{figure}

For
example, the states $\ket{111000}$ and $\ket{110010}$, which have Hamming
distance 2, differ by a single excitation. Similarly,
the states $\ket{011010}$ and $\ket{010101}$ differ by a double excitation. This connection between states and excitations is illustrated in Fig.~\ref{fig:excitation_links}.

A $U(2)$ rotation in the subspace spanned by the $k$-particle states $\ket{x},\ket{y}$ is therefore equivalent to a unitary performing the transformation
\begin{align}
U\ket{x} &= a\ket{x} + b\ket{y},\\
U\ket{y} &= d\ket{y} + c\ket{x},
\end{align}
while leaving every other basis state unchanged. For the specific case of states
with $n=2\ell$ qubits, this is accomplished by a unitary Givens rotation
$G^{(\ell)}$, as in Eqs.~\eqref{eq:g^l_1} and \eqref{eq:g^l_2}. When applied to
states with $n>2\ell$ qubits, the gate $G^{(\ell)}$ acts non-trivially on any
states where the $2\ell$ qubits in question are
set to $\ket{x}$ or $\ket{y}$, regardless of the state of the remaining
qubits. For instance, if $\ket{z}$ is a basis state of $m=n-2\ell$ qubits, it holds that 
\beq
G^{(\ell)}\ket{z}\ket{x}=a\ket{z}\ket{x}+b\ket{z}\ket{y},
\eeq
for all $z$. To address this issue, we can simply apply $G^{(\ell)}$ \emph{controlled} on the state of the remaining $n-2\ell$ qubits. We use the notation $C^{(m)}G^{(\ell)}$ to denote a $G^{(\ell)}$ gate controlled on the state of $m$ qubits. Controlling on the remaining qubits being in state $\ket{z^*}$ and defining $\ket{x'}=\ket{z^*}\ket{x}$ and $\ket{y'}=\ket{z^*}\ket{y}$, we then have that
\begin{align}
C^{(m)}G^{(\ell)}\ket{x'} &= a\ket{x'} + b\ket{y'},\\
C^{(m)}G^{(\ell)}\ket{y'} &= d\ket{y'} + c\ket{x'},
\end{align}
while leaving all other basis states unchanged. This is the desired two-dimensional transformation.

\begin{figure}[t]
\centering
\includegraphics[width=0.85\columnwidth]{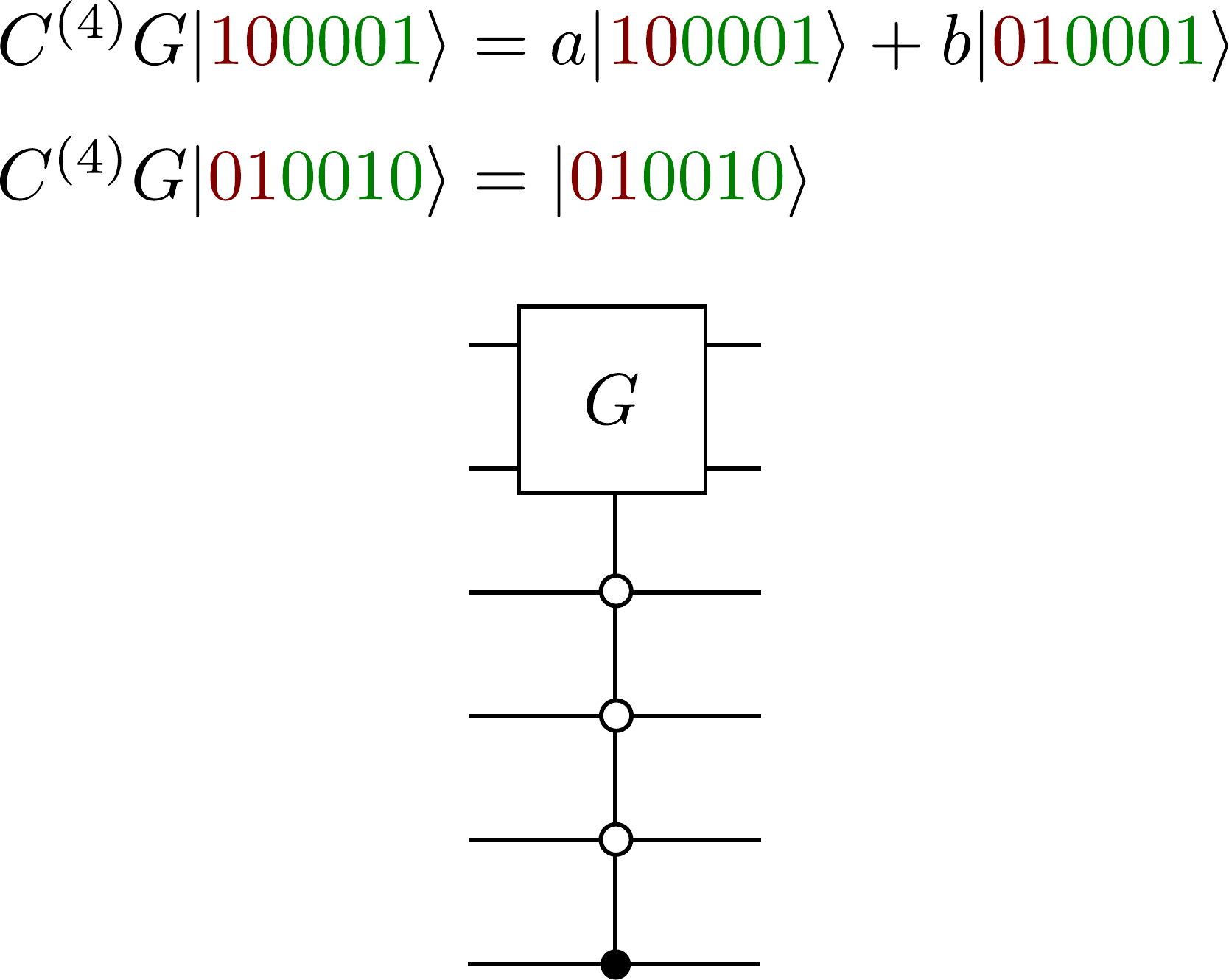}
\caption{Circuit diagram of a
  multi-controlled single-excitation gate. A single-excitation gate $G$ can act non-trivially on many states. To
  ensure that the desired rotation happens only between two target states, we
  can apply a multi-controlled gate $C^{(4)}G$ that acts as the identity on
  non-target states. In this example, the goal is to perform a rotation in the
  subspace of states $\ket{100001}$, $\ket{010001}$. This can be achieved by
  applying a single-excitation gate to the first two qubits, controlled on the
  state of the last four being $\ket{0001}$. This guarantees that other states,
  such as $\ket{010010}$, are left unchanged.}\label{fig:need_controls}
\end{figure}

Consider the states $\ket{100011}$ and $\ket{010011}$. They differ by an excitation from the first to the second qubit, and coincide on the remaining four qubits. A non-controlled single-excitation gate acting on the first two qubits would also perform a transformation on other subspaces, for example the one spanned by $\ket{101101}$ and $\ket{011100}$. However, controlling on the last four qubits being in state $\ket{0011}$ ensures that the gate $C^{(4)}G$ acts non-trivially only on the target two-dimensional subspace. The role of multiple controls is shown in Fig.~\ref{fig:need_controls}.

Overall, we conclude that any two-level $U(2)$ gate on the subspace of $k$-particle states on $n$ qubits can be implemented in terms of multi-controlled excitation gates $C^{(m)}G^{(\ell)}$, where $n=2\ell+m$. Following standard results, this implies that multi-controlled excitation gates are universal for particle-conserving operations.

\subsection{Single-excitation gates with multiple controls}

We now show that multi-controlled excitation gates can be decomposed in terms of multi-controlled single-excitation gates $C^{(m)}G$. The construction follows a similar approach to Ref.~\cite{nielsen2000quantum}. Suppose we wish to decompose a $U(2)$ Givens rotation on the subspace spanned by $\ket{x}, \ket{y}$, where the states have Hamming distance $2\ell$. The goal is to employ single-excitation gates to perform a permutation of all basis states such that the permuted versions of $\ket{x}$ and $\ket{y}$ differ by a single excitation. This can be achieved following similar principles to the construction of Gray codes. A controlled single-excitation gate can then be applied, followed by a reversal of the permutation.

\begin{figure}[t]
\centering
\includegraphics[width=0.85\columnwidth]{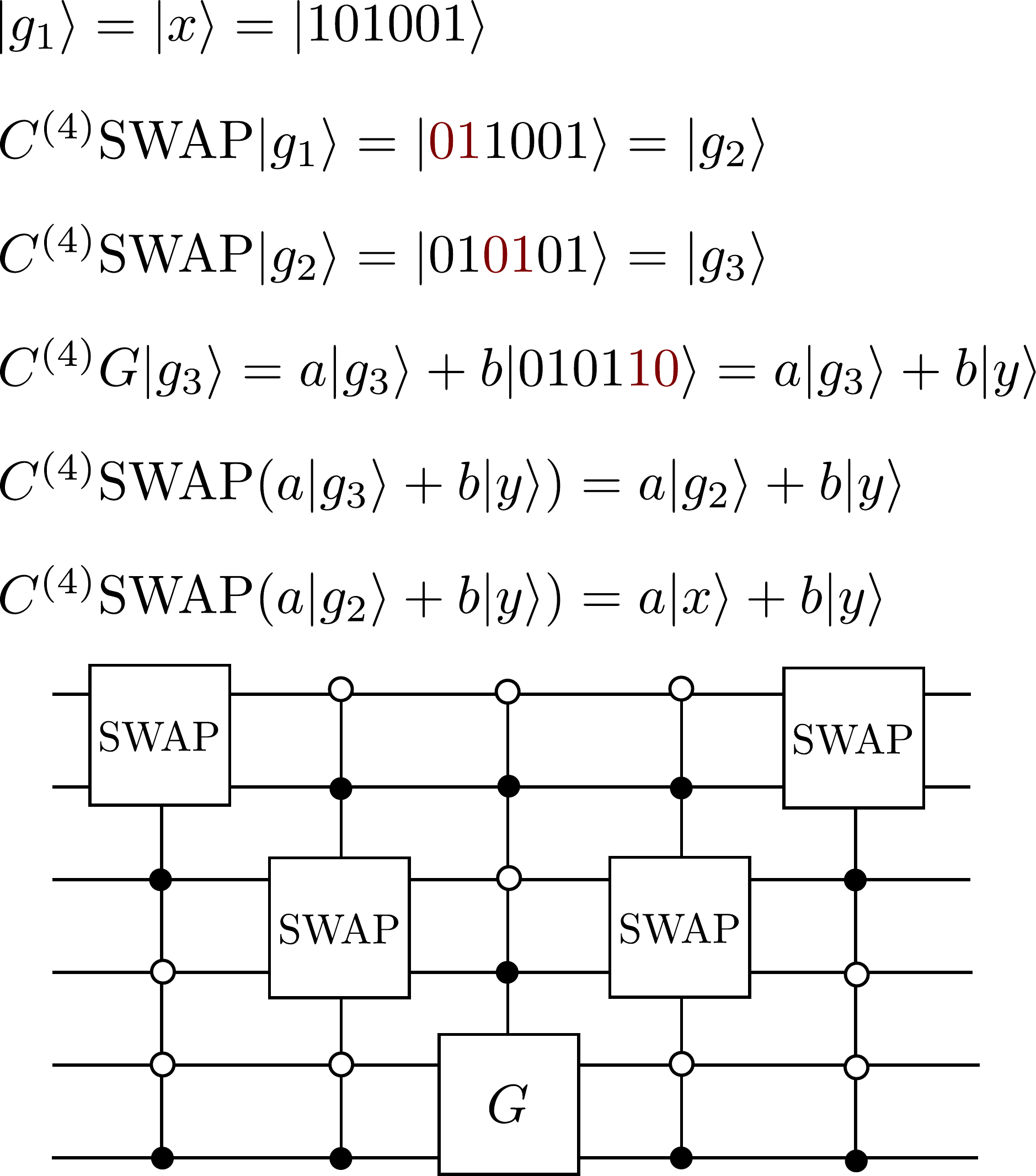}
\caption{Method for connecting any pair of states by a sequence of particle-conserving multi-controlled SWAP gates. The initial state $\ket{x}=\ket{101001}$ differs from the target state $\ket{y}=\ket{010110}$ by a triple excitation, which can be decomposed in terms of single excitations using a sequence of intermediary states. A circuit implementing a triple-excitation rotation can then be decomposed in terms of (i) multi-controlled SWAP gates performing a permutation of states, (ii) a multi-controlled single-excitation gate, and (iii) a reversal of the permutation. The end result is to create an arbitrary superposition $a\ket{x}+b\ket{y}$ of the target states. }\label{fig:single_ladder}
\end{figure}

For example, the states $\ket{101001}$ and $\ket{010110}$, which differ by a triple excitation, can be linked through the sequence 
$\ket{101001}\rightarrow \ket{011001}\rightarrow \ket{010101}\rightarrow \ket{010110}$, where each new state differs from the previous one by a single excitation. Each of the states in this sequence can be obtained from the previous one by applying a SWAP gate
\beq
\text{SWAP} = \begin{pmatrix}
1 & 0 & 0 & 0\\
0 & 0 & 1 & 0\\
0 & 1 & 0 & 0\\
0 & 0 & 0 & 1
\end{pmatrix},
\eeq
controlled on the state of all remaining qubits, i.e., by applying a $C^{(m)}\text{SWAP}$ gate. The SWAP gate is a special case of a single-excitation gate. As before, the control is required to ensure that the resulting permutation happens non-trivially only on the two-dimensional target subspace. This procedure is illustrated in Fig.~\ref{fig:single_ladder}.

We now describe the method in more detail. Without loss of generality, suppose that the states $\ket{x}$ and $\ket{y}$ differ on the first $2\ell$ qubits. The first step is to outline an ordered sequence of computational basis states $\ket{g_1}, \ket{g_2}, \ldots, \ket{g_{\ell+1}}$ such that all $\ket{g_i}, \ket{g_{i+1}}$ differ by a single excitation, and where $\ket{x}=\ket{g_1}$ and $\ket{y}=\ket{g_{\ell+1}}$. To build the circuit implementing the decomposition, we perform the following steps.
\begin{enumerate}
\item Apply a SWAP gate to the qubits where $\ket{x}=\ket{g_1}$ and $\ket{g_2}$ differ, controlled on all other qubits. This has the effect of swapping $\ket{g_1},\ket{g_2}$, while leaving all other states unchanged.
\item Follow the same procedure to swap $\ket{g_2}, \ket{g_3}$,  then $\ket{g_3}, \ket{g_4}$, and all other states until the final swap between $\ket{g_{\ell-1}}, \ket{g_{\ell}}$. This sequence of operations has the effect of mapping $\ket{x}\rightarrow \ket{g_{\ell}}$ while $\ket{g_{\ell+1}}=\ket{y}$ is left unchanged. 
\item Since $\ket{g_\ell}, \ket{g_{\ell+1}}$ differ by a single excitation, we perform a multi-controlled single-excitation gate that acts only on the subspace spanned by these states.
\item The circuit is completed by reverting all the swaps such that the resulting transformation is a rotation in the subspace spanned by $\ket{x}$ and $\ket{y}$, as desired.
\end{enumerate}

\subsection{Controlled single-excitation gates are universal}
\begin{figure}[t]
\centering
\includegraphics[width=0.93\columnwidth]{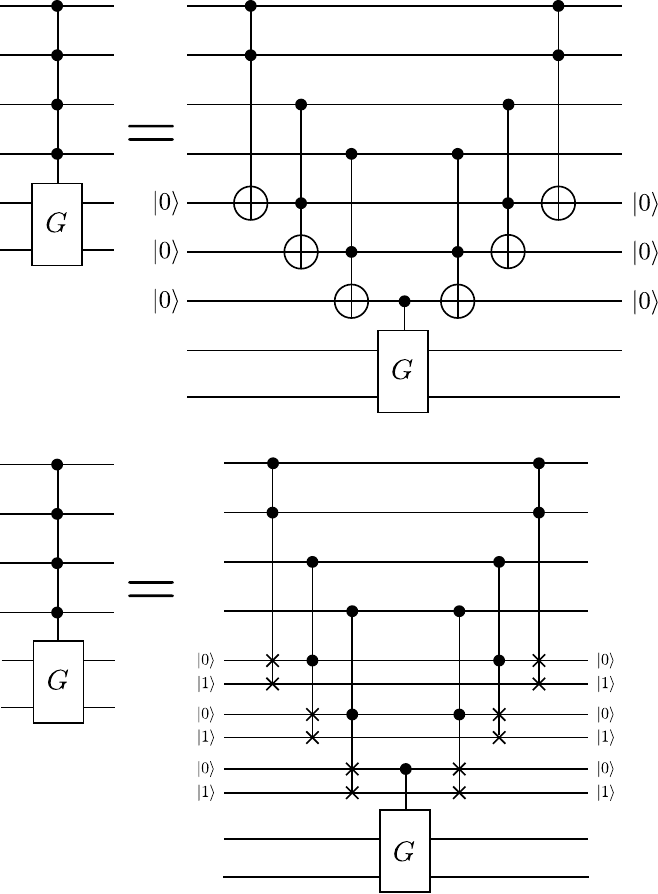}
\caption{A multi-controlled single-excitation gate can be decomposed in
  terms of a controlled single-excitation gate using a cascade of Toffoli gates. The Toffoli gates can be decomposed in terms of the particle-conserving
  controlled-controlled-SWAP gate using additional dual-rail qubits.}\label{fig:cascade}
\end{figure}

Given a gate controlled on a single qubit, there exist well-established methods to extend the control to additional qubits~\cite{nielsen2000quantum}. Suppose we want to control the operation on the state of $m$ qubits $\ket{z_1z_2\cdots z_m}$. The strategy relies on employing $m-1$ auxiliary qubits and Toffoli gates (controlled-CNOT gates), as shown in Fig.~\ref{fig:cascade}. 

\begin{figure}[t]
\centering
\includegraphics[width=0.8\columnwidth]{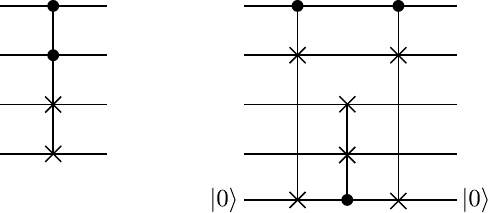}
\caption{A controlled-controlled-SWAP gate can be decomposed in terms of three Fredkin gates using an auxiliary qubit~\cite{aaronson2015classification}.}\label{fig:ccswap}
\end{figure}

Toffoli gates are not particle-conserving, but they can be decomposed in terms of particle-conserving Fredkin gates by
replacing the auxiliary qubits with dual-rail qubits $\ket{\tilde{0}}:=\ket{01}$,
$\ket{\tilde{1}}:=\ket{10}$. In this case a Toffoli gate is equivalent to a
controlled-controlled-SWAP gate, since swapping $\ket{01}$ and $\ket{10}$
applies a NOT gate to the dual-rail qubit. As shown in Ref.~\cite{aaronson2015classification}, the controlled-controlled-SWAP gate
can then be decomposed into three Fredkin (controlled-SWAP) gates with the help of an auxiliary qubit, as shown in Fig.~\ref{fig:ccswap}.

Overall, we have shown that single-excitation gates controlled on multiple qubits, which were previously shown to be universal, can be decomposed into controlled single-excitation gates, which are therefore also universal. This concludes the proof that controlled single-excitation gates are universal for particle-conserving unitaries.

\section{State preparation}\label{sec:state-prep}

Universal gate sets for particle-conserving unitaries can also be used to prepare arbitrary states of a fixed number of particles. Here we discuss how controlled single-excitation gates can be used for this purpose. We follow the strategy of Ref.~\cite{mottonen2005transformation}.

Consider a system of $k$ particles on $n$ qubits, spanning a space of dimension $d=\binom{n}{k}$. Any such state can be written as $\ket{\psi}=\sum_{x}c_x \ket{x}$,
where the sum is over all $n$-bit strings $x$ of Hamming weight $k$. As shown before, an arbitrary $U(2)$ rotation in the subspace of any pair of states $\ket{x}, \ket{y}$ can be performed by a suitable decomposition into controlled single-excitation gates.

Consider a lexicographical labelling of all bit strings of Hamming weight $k$ as $\ket{x_1}, \ket{x_2}, \ldots, \ket{x_{d}}$. For instance, in the case of $n=3$ and $k=2$ we have $\ket{x_1}=\ket{011}$, $\ket{x_2}=\ket{101}$, and $\ket{x_3}=\ket{110}$. An arbitrary state can then be written as 
\beq\label{eq:arbitrary_state}
\ket{\psi}=\sum_{i=1}^dc_i \ket{x_i}.
\eeq
We describe a method to prepare any such state starting from the reference state $\ket{x_1}$.

First, apply the multi-controlled excitation operation in the subspace $\ket{x_1}, \ket{x_2}$ that performs the mapping
\beq
\ket{x_1}\rightarrow \alpha_1 \ket{x_1} + c_2 \ket{x_2},
\eeq
where $\alpha_1=\sqrt{1-|c_2|^2}$. Then, apply the multi-controlled excitation operation in the subspace $\ket{x_1}, \ket{x_3}$. This performs the mapping
\beq
\alpha_1 \ket{x_1} + c_2 \ket{x_2}\rightarrow \alpha_1\alpha_2 \ket{x_1} + c_2 \ket{x_2} + \alpha_1 c'_3\ket{x_3},
\eeq
where we set $c'_3=c_3/\alpha_1$ and $\alpha_2=\sqrt{1-|c'_3|^2}$. This ensures that the coefficient in front of $\ket{x_3}$ is precisely the desired one, $c_3$. This process can be repeated for each of the remaining states $\ket{x_4},\ldots\ket{x_d}$. The result is to prepare the state
\beq
\ket{\psi}=\alpha\ket{x_1}+\sum_{i=2}^{d}c_i\ket{x_i},
\eeq 
where $\alpha=\prod_{i=1}^{d-1}\alpha_i$. This state is normalized, which from Eq.~\eqref{eq:arbitrary_state} implies that $|\alpha|^2=|c_1|^2$. To ensure that in fact $\alpha=c_1=|c_1|e^{i\theta}$, it suffices to choose $\alpha_1, \alpha_2,\ldots, \alpha_{d-2}$ to be positive real numbers, as we have done, and set $\alpha_{d-1}=|\alpha_{d-1}|e^{i\theta}$ to prepare the desired state. Note that for a unitary $U=\begin{pmatrix} a & c\\ b & d\end{pmatrix}$, $a$ can be guaranteed to be real for any $c$ by choosing 
\begin{align*}
d&=-\sqrt{1-|c|^2}\\
b^*&=\frac{c\sqrt{1-|c|^2}}{a}.
\end{align*}

In practice, it is possible to simplify this general strategy when applied to particular cases. Since the gates act only on a specific superposition of basis states, controls only need to be applied on qubits where the states in the superposition differ. This is useful if the target state has support only on a specific subspace. For example, the first excitation gate performing the mapping $\ket{x_1}\rightarrow \alpha_1 \ket{x_1} + c_2 \ket{x_2}$ does not need to be controlled. Furthermore, excitation gates can be chosen to act on different reference states in order to create new superpositions.  

Consider the six-qubit state
\beq
c_1\ket{110000} + c_2\ket{001100}+ c_3\ket{000011}+ c_4\ket{100100},
\eeq
which corresponds to a superposition of the four basis states that contribute most significantly to the ground-state energy of the $\text{H}^+_3$ molecule in a minimal basis set. The state can be prepared as follows.

\begin{figure}[t]
\centering
\includegraphics[width=0.7\columnwidth]{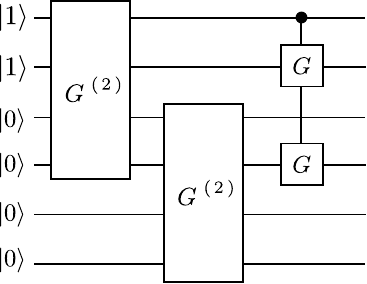}
\caption{A quantum circuit for preparing the state $c_1\ket{110000} + c_2\ket{001100}+ c_3\ket{000011}+ c_4\ket{101000}$ for arbitrary values of the coefficients $c_1, c_2, c_3, c_4$. Double-excitation gates are applied to the first four qubits and then to the last four qubits. Finally, a controlled single-excitation gate is applied to qubits 2 and 4, controlled on the state of the first qubit.}\label{fig:state_prep}
\end{figure}

Starting from $\ket{110000}$, apply a double-excitation gate to the first four qubits to prepare the state $a_1\ket{110000} + c_2\ket{001100}$. This does not need to be controlled on any qubit. Then apply a double-excitation gate to qubits 3,4,5,6 to prepare the state $a_1\ket{110000} + c_2\ket{001100} + c_3\ket{000011}$, where we're using $\ket{001100}$ as the reference. This again does not need to be controlled. To obtain the desired state, apply a single-excitation gate to qubits 2 and 4, controlled on the first qubit being in state $\ket{1}$, which prevents mixing with the state $\ket{001100}$. This construction is shown in Fig.~\ref{fig:state_prep}.

\section{Variational quantum circuits}\label{sec:variational}

\begin{figure}[t]
\centering
\includegraphics[width=0.6\columnwidth]{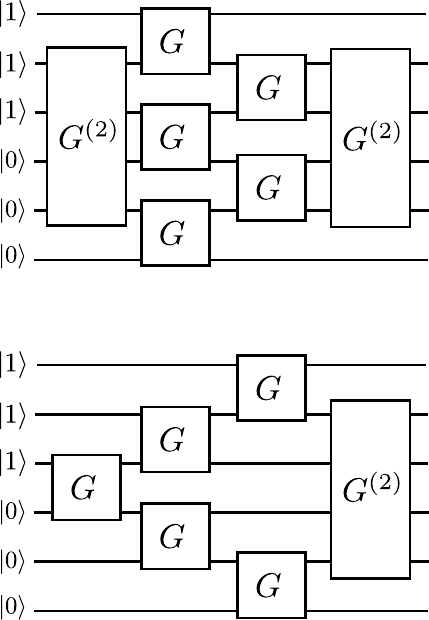}
\caption{Example circuit architectures constructed from Givens rotations as fundamental building blocks. Since these operations are particle-conserving, it is possible to compose them arbitrarily to create various types of particle-conserving circuits. Excitation gates without controls are used to access larger subspaces with fewer gates, whose parameters may then be optimized for specific purposes.}\label{fig:variational}
\end{figure}

We discuss implications of our results for variational quantum circuits. Our
universality result suggests the use of controlled single-excitation gates as
building blocks for variational quantum circuits. For example, the state
preparation algorithm described above can be employed as a template where the
rotation angles for each gate are free parameters of the model. In this context,
multiple controls are not necessary; instead, by employing uncontrolled
excitation gates it is possible to reach a larger subspace of states using fewer
gates, but generally this makes it more challenging to prepare specific target states. In the example of Fig.~\ref{fig:state_prep}, dropping the control on the last
single excitation gate leads to a state with a non-zero coefficient on the
additional basis state $\ket{011000}$. Examples of variational circuits designed using Givens rotations as building blocks are shown in Fig.~\ref{fig:variational}. Below we describe two specific strategies for building quantum circuits for quantum chemistry applications, and derive analytical gradient formulas for Givens rotations.

\subsection{All singles and doubles}

In the context of quantum computing, the unitary coupled-cluster singles and doubles (UCCSD) ansatz is often expressed in terms of fermionic operators, which are then mapped to complicated qubit gates. In similar spirit to the qubit-coupled-cluster approach of Ref.~\cite{ryabinkin2018qubit}, we can instead consider a circuit where single and double excitations are respectively implemented using Givens rotations $G$ and $G^{(2)}$. A quantum circuit can then be defined consisting of all possible single and double excitation gates that act non-trivially on the reference state without flipping the spin of the excited particles. The resulting circuit is analogous to a Trotterized implementation of UCCSD to first level, but where all gates are Givens rotations. This is illustrated in Fig.~\ref{fig:circuits}.

\subsection{Adaptive circuits}

Adaptive strategies such as those presented in Refs.~\cite{grimsley2019adaptive, tang2021qubit} can be implemented by selecting Givens rotations instead of fermionic excitations in the constructions. The main idea is that instead of designing circuits that work well for all molecules, we can instead build specific circuits that are custom-built for each molecule. Hence, what is general is the method for building custom circuits, not the circuits themselves.

\begin{figure}[t]
\centering
\includegraphics[width=\columnwidth]{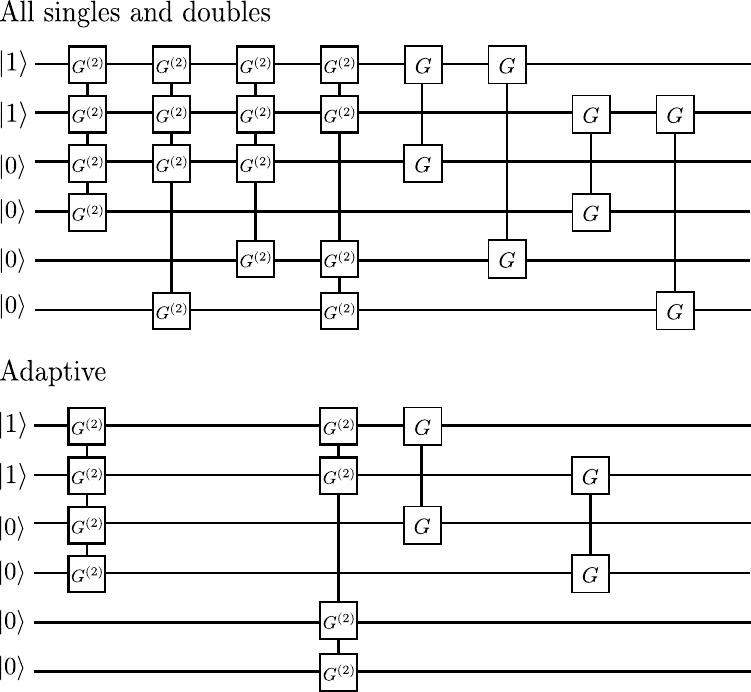}
\caption{Strategies for building quantum circuits for quantum chemistry. The first strategy consists of selecting all single and double excitation gates that do not flip the spin of the particles. In an adaptive strategy, after initializing gate parameters, we compute the gradient for each gate and keep only those that are above a given threshold.}\label{fig:circuits}
\end{figure}

A simple yet powerful strategy is to build a circuit consisting of all double and single excitation gates, randomly initialize all parameters, and compute the gradient for each gate. The final circuit is constructed by keeping only those gates such that the norm of their gradient exceeds a fixed threshold. This is shown in Fig.~\ref{fig:circuits}.

\subsection{Analytical gradients}
We derive analytic gradient formulas for Givens rotations. If $\tilde{H}$ is the generator of a unitary $\tilde{U}(\theta)=e^{i\tilde{H}\theta}$, then the generator of the unitary $U(\theta)=\id\oplus \tilde{U}(\theta)$ is $H=\bm{0}\oplus \tilde{H}$, where $\bm{0}$ denotes the zero operator. As shown in~\cite{kottmann2021feasible}, generators of this form can be decomposed as
\begin{align}
H &= \frac{1}{2}(H_+ + H_-),\\
H_{\pm} &=(\pm \id)\oplus \tilde{H}.
\end{align}
We can then write
\beq
U(\theta)=e^{i\theta G_+/2}e^{i\theta G_-/2},
\eeq
and define the gates
\beq
U_{\pm}(\theta)=e^{i\theta G_{\pm}}.
\eeq
The operators $H_\pm$ satisfy
\begin{align}
H_\pm^2 &= \id,\\
[H_+, H_-] &= 0.
\end{align}

As shown in~\cite{mari2021estimating}, any unitary $U(\theta)$ with generator that is self-inverse satisfies the parameter-shift rule
\begin{align}
\frac{\partial C(\theta)}{\partial \theta}=\frac{C(\theta+s)-C(\theta-s)}{2\sin(s)},
\end{align}
for any cost function that can be written as $C(\theta)=\langle\psi|U^\dagger(\theta)KU(\theta)|\psi\rangle$, where $K$ is an observable.

This parameter-shift rule applies to the gates $U_{\pm}(\theta)$, which means that derivatives of $U(\theta)$ can be obtained by writing
$U(\theta)=U_+(\theta/2)U_-(\theta/2)$ and computing derivatives of the gates $U_{\pm}(\theta)$. This technique can be employed for any Givens rotation whose generator $\tilde{H}$ is self-inverse. For example, in the case of the Givens rotation of Eq.~\eqref{eq:givens}, we can write $G(\theta)=G_{+}(\theta/2)G_{-}(\theta/2)$ where the unitaries
\beq
G_{\pm}(\theta)=
\begin{pmatrix}
e^{\pm i\theta} & 0 & 0 & 0\\
0 & \cos \theta & -\sin \theta & 0\\
0 & \sin\theta & \cos \theta & 0\\
0 & 0 & 0 & e^{\pm i\theta}
\end{pmatrix},
\eeq
satisfy the parameter-shift rule.

\section{Conclusion}\label{sec:conclusion}
\vspace{-1mm}
We have shown that controlled single-excitation gates are universal for particle-conserving unitaries. These three-qubit gates are Givens rotations performing a transformation in a two-dimensional subspace of states $\ket{01}, \ket{10}$, controlled on the state of a third qubit. The states $\ket{01}, \ket{10}$ can be interpreted as a dual-rail encoding of a single qubit, thus making controlled single-excitation gates analogous to controlled single-qubit gates. This provides a unified approach for combining techniques from quantum computing into a form that is directly applicable to quantum chemistry: instead of single-qubit gates and controlled two-qubit gates, we can focus on single-excitation Givens rotations and controlled Givens rotations. This connection could potentially open avenues for more efficient compilation strategies that employ Givens rotations as the universal gate set while employing standard techniques applicable for general circuits.

The proof of universality relies on the ability to control excitation gates on multiple qubits. This leads to decompositions employing auxiliary qubits and controlled single-excitation gates. These constructions are helpful proof techniques, but may not be the optimal approaches to compiling circuits. Instead, quantum circuits can be designed using Givens rotations directly. The fact that controlled Givens rotations are universal and can be freely composed, while satisfying the main symmetries of fermionic systems, allows for a versatile approach to quantum algorithm design. For example, Givens rotations can also be used to directly enforce spin conservation by coupling only spin-orbitals with equal spin projection. More generally, they can be composed to perform arbitrary transformations on any desired subspace of the fixed-particle-number manifold. For all such circuits, we can obtain analytical gradient rules that can be used to optimize circuits when desired. Overall, our results indicate that Givens rotations are better abstractions than fermionic excitation gates.

For quantum chemistry applications, it is likely that custom algorithms will be needed to tackle specific problems and molecules. Instead of preparing a menu of quantum circuits and algorithms for each instance, we aim to provide scientists with a set of universal ingredients that can be employed to craft tailored solutions. Our results serve as a unifying framework for quantum computational chemistry where every algorithm is a unique recipe built from the same universal ingredients: Givens rotations. 

\appendix

\begin{figure*}[t!]
\centering
\includegraphics[width=0.95\textwidth]{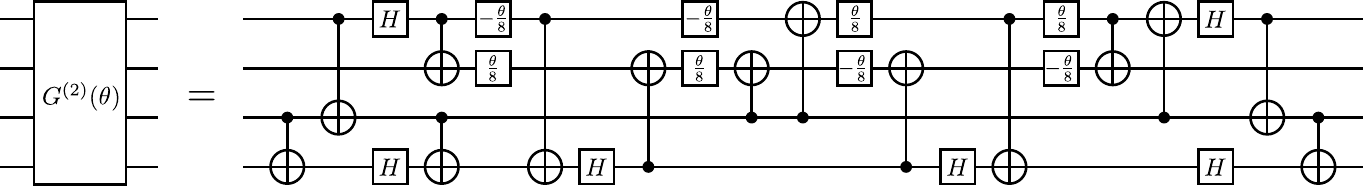}
\caption{Decomposition of a double excitation gate into single-qubit rotations
  and CNOTs. All gates denoted by $\pm\frac{\theta}{8}$ are Pauli $Y$ rotations. Adapted from \cite{anselmetti2021local}.}\label{fig:double_ex_decomposition}
\end{figure*}

\begin{figure}
\centering
\includegraphics[width=0.95\columnwidth]{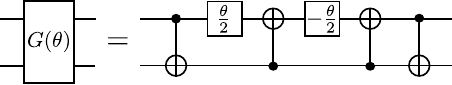}
\caption{Decomposition of a single excitation gate into single-qubit rotations
  and CNOTs. The gates denoted by $\pm\frac{\theta}{2}$ are Pauli $Y$ rotations. The middle four gates constitute a controlled $Y$ rotation.}\label{fig:single_ex_decomposition}
\end{figure} 

\section{Decompositions of excitation operators}

While controlled single excitation operators comprise a universal gate set,
hardware constraints often require a decomposition of these operators over
the gate set of single-qubit rotations and CNOTs.
Fig.~\ref{fig:single_ex_decomposition} presents such a decomposition for a single-excitation gate. It is straightforward to extend this to a controlled
version by applying a control to each individual gate. 
This would produce Toffolis and controlled-$Y$
rotations, both of which can be further decomposed over the gate set.
Fig.~\ref{fig:double_ex_decomposition} presents a decomposition for the
four-qubit double-excitation gate. This decomposition was adapted from that of Ref.~\cite{anselmetti2021local}.

\bibliographystyle{quantum}
\bibliography{refs-doi}

\end{document}